\documentclass[12pt]{article}
\usepackage{geometry}
\usepackage{hyperref}
\usepackage{authblk} 
\usepackage{microtype}
\usepackage{titlesec}
\usepackage{amsbsy} 
\usepackage{amssymb}
\usepackage{amsmath}
\usepackage{bm}
\usepackage{array}
\usepackage{booktabs}
\usepackage{adjustbox}
\usepackage{multirow}
\usepackage{arydshln}
\usepackage{multicol}
\usepackage{pdfpages}
\usepackage{graphicx}
\usepackage{caption}
\usepackage{subcaption}
\usepackage{natbib}
\usepackage{enumitem}
\usepackage{glossaries}
\usepackage{setspace}
\usepackage{lineno} 

\newcommand{\bigzero}{\mbox{\normalfont\Large 0}}
\titleformat{\paragraph}
{\normalfont\it\normalsize}{\theparagraph}{1em}{}
\titlespacing*{\paragraph}
{0pt}{3.25ex plus 1ex minus .2ex}{1.5ex plus .2ex}

\begin{document}

\title{Overview and Introduction to Development of Non-Ergodic Earthquake Ground-Motion Models \\~
       [Published in Bulletin of Earthquake Engineering]}

\author[1]{Grigorios Lavrentiadis\thanks{glavrentiadis@ucla.edu}}
\author[2]{Norman A. Abrahamson}
\author[1]{Kuehn M. Nicolas}
\author[1]{Yousef Bozorgnia}
\author[3]{Christine A. Goulet}
\author[4]{Anže Babič}
\author[5]{Jorge Macedo}
\author[4]{Matjaž Dolšek}
\author[6]{Nicholas Gregor}
\author[7]{Albert R. Kottke}
\author[2]{Maxime Lacour}
\author[5]{Chenying Liu}
\author[3]{Xiaofeng Meng}
\author[2]{Van-Bang Phung}
\author[2]{Chih-Hsuan Sung}
\author[8]{Melanie Walling}

\affil[1]{Natural Hazards Risk and Resiliency Research Center (NHR3), University of California, Los Angeles}
\affil[2]{Department of Civil Engineering, University of California, Berkeley}
\affil[3]{Southern California Earthquake Center, University of Southern California}
\affil[4]{Faculty of Civil and Geodetic Engineering, University of Ljubljana}
\affil[5]{School of Civil and Environmental Engineering, Georgia Institute of Technology}
\affil[6]{Consultant}
\affil[7]{Geosciences, Pacific Gas \& Electric Co.}
\affil[8]{Performance-Based Risk Assessment, GeoEngineers, Inc.}
\renewcommand\Authands{ and }

\maketitle

\begin{abstract}

This paper provides an overview and introduction to the development of non-ergodic ground-motion models, GMMs.
It is intended for a reader who is familiar with the standard approach for developing ergodic GMMs.
It starts with a brief summary of the development of ergodic GMMs and then describes different methods that are used in the development of non-ergodic GMMs with an emphasis on Gaussian Process (GP) regression, as that is currently the method preferred by most researchers contributing to this special issue.
Non-ergodic modeling requires the definition of locations for the source and site  characterizing the systematic source and site effects; the non-ergodic domain is divided into cells for describing the systematic path effects.  
Modeling the cell-specific anelastic attenuation as a GP, and considerations on constraints for extrapolation of the non-ergodic GMMs are also discussed.
An updated unifying notation for non-ergodic GMMs is also presented, which  has been adopted by the authors of this issue.

\end{abstract}

\section{Introduction} \label{sec:intro}

Due to the limited number of ground-motion recordings in a region, the traditional approach to developing ground-motion models (GMMs) for use in probabilistic seismic hazard analysis (PSHA) has been to combine data from similar tectonic environments around the world together and develop a model for the scaling with magnitude, distance, and site conditions.
The median and aleatory variability of a GMM are assumed to be applicable to any location within the broad tectonic category.
This is known as the ergodic assumption in ground-motion modeling \citep{Anderson1999}.

The traditional approach of developing ergodic GMMs leads to a stable global average of the ground motion for a given scenario, but a large aleatory variability between an observation and the global average. 
With the large increase in the number of ground-motion instruments and recordings over the last decade, it has become clear that there are significant systematic differences in ground motion based on the location of the site and the source.
As a result, the ergodic GMMs generally may not work well for a specific site/source location.
This has prompted the development of non-ergodic ground-motion models in which these location-specific effects are modeled explicitly, which reduces the aleatory variability. 
The uncertainty in the estimate of the site-specific effects is then part of the epistemic uncertainty.
As a general classification, uncertainties are treated as epistemic if they are expected to be reduced by gathering more data. Variabilities are treated as aleatory if the increase of data is not expected to systematically reduce their range \citep{DerKiureghian2009}.
 
\begin{figure}
    \centering
    \includegraphics[width = .45\textwidth]{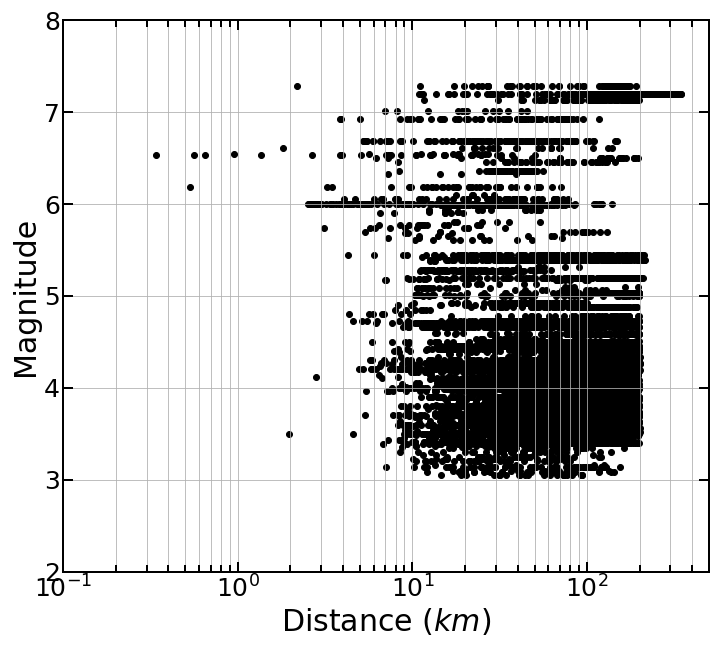}
    \caption{Magnitude-distance distribution of the California subset of the NGAWest2 dataset. }
    \label{fig:M-R_dist}
\end{figure}

An important difference between the application of statistics in GMMs as compared to most other fields is the use of constraints to ensure proper extrapolation. 
In other fields, the assumption is that the key behaviors are represented by the available data. 
Thus, the goal of statistics is to find the trends in the data.
However, the problem is more complicated in earthquake GMMs as they are often applied to earthquake scenarios outside the range that is well constrained by the data (i.e., it is an extrapolation problem).
For instance, Figure \ref{fig:M-R_dist} shows the magnitude-distance distribution of the California subset of the NGA-West2 database \citep{Ancheta2014}, which is often used for the development of GMMs for California. 
In this dataset, the magnitudes range from $3$ to $7.2$, with the majority of the events being between magnitude $3$ and $5$, and the distance ranges from $1$ to $400 km$, with the majority of the recordings being between $20$ and $200 km$. 
However, in PSHA, large-magnitude and short-distance scenarios often control the hazard.
For example, in the San Francisco Bay Area, it is common to have faults that are less than $10 km$ away from a site and are capable of producing larger than $M 7$ earthquakes.
It is the difference between the range of the scenarios that are used to derive a GMM and the range of scenarios on which a GMM is applied that makes the proper extrapolation of the ground motion an important aspect of a GMM. 

This paper serves as an overview and introduction to the development of non-ergodic GMMs, with an emphasis on the varying coefficient models (VCM) developed with Gaussian Processes (GPs) regression.
The combination of a VCM GMM with the cell-specific anelastic attenuation and considerations regarding the extrapolation of non-ergodic GMMs are also discussed. 
An updated notation for key elements of non-ergodic GMMs is also presented.

\section{Proposed Notation}

The proposed notation is intended to help the reader understand the role of different terms in a GMM and facilitate the comparison of the different non-ergodic models in this special issue. 

The model variables are categorized into two groups: the model parameters ($\vec{\theta}$) and model hyperparameters ($\vec{\theta}_{hyp}$). 
The $\vec{\theta}$ includes the ergodic and non-ergodic terms that directly affect the ground motion, while $\vec{\theta}_{hyp}$ includes the set of variables that control the behavior of the ergodic and non-ergodic terms and have an indirect effect on the ground motion. 
An example of a model parameter is the coefficient for the linear magnitude term, and an example of a hyperparameter is the between-event standard deviation.  

The ergodic coefficients of the GMM are denoted as $c_i$ where $i$ is the number of the ergodic term, and the non-ergodic coefficients are denoted as $c_{i,X}$ or $\delta c_{i,X}$.
The subscript $X$ (subscript after the comma) can be the letters $E$, $P$, or $S$ depending on whether the non-ergodic coefficient in question is intended to capture systematic effects related to the source (earthquake), path, or site.
The notation $\delta$ is used to differentiate between non-ergodic coefficients that have a zero mean and act as adjustments to ergodic coefficients, and stand-alone non-ergodic coefficients that encompass both the average scaling and the systematic effects.  
For instance, the non-ergodic term $\delta c_{1,E}$ acts on top of $c_1$ to capture the systematic effects related to the source.
Alternatively, the same behaviour can be modeled with $c_{1,E}$ which is equal to the sum of the ergodic coefficient and the non-ergodic adjustment ($c_{1,E} = c_1 + \delta c_{1,E}$). 
The non-ergodic adjustments, $\delta c_{i,X}$, are typically used when a non-ergodic GMM is developed with an ergodic GMM as a "backbone" model, in which case the non-ergodic adjustments are estimated based on the total ergodic residuals. For example, \cite{Kuehn2019} and \cite{Lavrentiadis2021} are non-ergodic GMMs derived with this approach. 
The non-ergodic coefficients, $c_{i,X}$, are used when a non-ergodic GMM is directly estimated with the log of the ground motion as a response variable as in the case of \cite{Landwehr2016}.

Many of the non-ergodic GMMs in this issue used the cell-specific anelastic attenuation, first proposed by \cite{Dawood2013}, to model the systematic path effects. 
In the proposed notation, the vector of attenuation coefficients of all the cells is denoted as $\vec{c}_{ca,P}$, and cell path segments are denoted as $\Delta \vec{R}$. 
The total anelastic attenuation is equal to $\vec{c}_{ca,P} \cdot \Delta \vec{R}$.

The terms $\delta L2L$, $\delta P2P$, and $\delta S2S$ introduced by \cite{AlAtik2010} are used here to describe the total non-ergodic effects related to the source (they used L for location), path, and site, respectively. 
For instance, if the constant $c_1$ is modified by two site adjustments, $\delta c_{1a,S}$ and $\delta c_{1b,S}$, to express the systematic site effects, then $\delta S2S$ is equal to $\delta c_{1a,S} + \delta c_{1b,S}$.
Similarly, if the non-ergodic adjustment to the geometrical spreading coefficient, $\delta c_{3,P}$, and the cell-specific anelastic attenuation, $\vec{c}_{ca,P}$ are used to express the systematic path effects, then $\delta P2P = \ln({R}) \delta c_{3,P} + \vec{c}_{ca,P} \cdot \Delta \vec{R} - c_7 R$, where, in this example, $c_7$ is the ergodic anelastic attenuation coefficient. 
The $c_7 R$ term is subtracted from the median prediction from the GMM to remove the systematic effects that are included in the cell-specific anelastic attenuation.

The scale and correlation length which control the spatial distribution of the non-ergodic terms are denoted as $\omega_{i,X}$ and $\ell_{i,X}$. 
An in-depth discussion on modeling the non-ergodic terms as GPs, where $\omega_{i,X}$ and $\ell_{i,X}$ are defined, is provided in section \ref{sec:gp}.
In keeping with the \cite{AlAtik2010} notation, the total epistemic uncertainty of the non-ergodic source, path, and site effects are denoted as $\tau_{L2L}$, $\phi_{P2P}$, and $\phi_{S2S}$.
Expanding from the example above, if $\omega_{1a,S}$ and $\omega_{1b,S}$ correspond to the scales of $\delta c_{1a,S}$ and $\delta c_{1b,S}$, then the epistemic uncertainty of the site effects is $\phi^2_{S2S} = \omega_{1a,S}^2 + \omega_{1b,S}^2 $.

The response variable of the regression is denoted as $y$.
For a pseudo-spectral acceleration ($PSA$) or Effective Amplitude Spectrum ($EAS$) GMM, $y$ is equal to $\ln(PSA)$ or $\ln(EAS)$.

The location of the source, site, etc. are required in the non-ergodic GMMs included in this issue to define the spatially-varying non-ergodic terms.
The coordinates of the earthquake, site, and mid-point between the source and site are denoted as: $t_E$, $t_S$, $t_{MP}$, respectively.
The definition of the location of the earthquake, $t_E$ (e.g. epicenter, the closest point to the site, etc.) is defined in each study. 
Similarly, the cell coordinates are denoted as $t_C$; the exact point of the cell (e.g. center, lower left corner, etc.) to which $t_C$ corresponds is defined in each study. 

The star superscript is used to denote the new scenarios and values of non-ergodic coefficients predicted for the new scenarios. For instance, $t^*_E$ corresponds to the source locations of the new scenarios where systematic source effects will be predicted, and $\delta c^*_{i,E}(t^*_E)$ corresponds to the non-ergodic source adjustments of these scenarios.  

A list of abbreviations and a glossary of all terms used in this special issue are provided at the end of this paper.

\section{Development of Ergodic Ground-Motion Models} \label{sec:erg_gmm}

A typical GMM has a model for the base magnitude, distance, and linear site scaling, and may include more complicated features for non-linear site response, hanging-wall effects, basin effects, etc.
For example, the median for the ASK14 \citep{Abrahamson2014} GMM has the following form:
\begin{equation} \label{eq:gmm_ask14}
\begin{aligned}
    f_{erg}(M,R_{rup},V_{S30}, ...) = & c_1 + c_2 M + c_3 (8.5-M)^2 + (c_4 + c_5 M) ln( R_{rup} + c_6) + c_7 R \\
                          & + c_8 F_{RV} + c_9 F_{N} + c_{10} \ln(V_{S30}/V_{ref})            \\
                          &  + f_{NL}(V_{S30}, z_{1100}) + f_{HW}(M,R_{rup},R_x,Dip) 
\end{aligned}
\end{equation}
where $f_{erg}$ is the function for the ergodic median ground-motion, $c_i$ are the ergodic scaling coefficients, $F_{RV}$ and $F_{N}$ are the reverse and normal fault scaling factors, $f_{NL}$ is the non-linear site effects scaling, $f_{HW}$ is the hanging wall scaling, $M$ is the moment magnitude, $R_{rup}$ is the closest distance to the rupture plane, $R_x$ is the horizontal distance from the top edge of the rupture measured perpendicular to the fault strike, $V_{S30}$ is the time-average shear wave velocity at the top $30 m$, $V_{ref}$ is the reference $V_{S30}$ for the linear site amplification, $z_1100$ is the depth to $1100 m/sec$ shear-wave velocity, and $Dip$ is the fault dip angle.  

A key aspect of GMMs used for seismic hazard studies is that in engineering applications, they need to be extrapolated 
outside the data range.
Although a GMM is developed through a regression analysis, constraints are often imposed on the coefficients to ensure that the GMM extrapolates consistently with a physical-based scaling. 

Because the recordings for a single earthquake are correlated, it is common to use a mixed-effects regression when developing GMMs:
\begin{equation}
    y_{es} = f(M,R_{rup},V_{S30}, ...) + \delta W_{es} + \delta B_e
\end{equation}
\noindent in which the left-hand-side is the observed ground motion ($y_{es}$), while the right-hand side includes $\delta B_e$ which is the between-event aleatory term for the $e^{th}$ earthquake and $\delta W_{es}$ which is the within-event aleatory term for the $s^{th}$ station and from the $e^{th}$ earthquake.  
$\delta B_e$ and $\delta W_{es}$ are assumed to be normally distributed with zero mean and $\tau$ and $\phi$ standard deviations, respectively.

\subsection{Maximum Likelihood Estimation} \label{sec:mle} 

In GMM development, the maximum likelihood estimation (MLE) is often used to obtain point estimates of the GMM coefficients (fixed terms) and standard deviations of the aleatory terms (random terms). 
In the past, the procedure outlined in \cite{Abrahamson1992} was commonly used to estimate the mixed-effect terms. 
More recently, statistical packages such as LME4 \citep{lme4} in the statistical software R \citep{R} are used to obtain point estimates and significance test statistics of the mixed terms. 

In MLE, the parameters of the model are estimated by maximizing the log-likelihood function, or more commonly by minimizing the negative of the log-likelihood function. 
The log-likelihood is a measure of how likely it is to observe the data given the model parameters. 
With the assumption that $\delta B_e$ and $\delta W_{es}$ are independent with no spatial correlation and that they are normally distributed, the log-likelihood function is given by:
\begin{equation} \label{eq:erg_Lfun}
    \ln \, \mathcal{L} = \frac{N}{2} \ln(2 \pi) - \frac{1}{2} \ln \, |\mathbf{C}| 
    - \frac{1}{2} (\vec{y} - \vec{\mu})^T 
    \mathbf{C}^{-1} (\vec{y} - \vec{\mu} )
\end{equation}

\noindent in which $\vec{y}$ is a vector with all ground motion observations, $\vec{\mu}$ is the median ground motion evaluated by $f_{erg}(M,R_{rup},...)$ for the parameters of all ground motion observations, and $\mathbf{C}$ is the covariance matrix which is given by:
\begin{equation} \label{eq:cov_mat_erg}
    \begin{aligned}
    \mathbf{C} =& \phi^2 \mathbf{I}_N + \tau^2 \Sigma_{i=1}^{N_e} \mathbf{1}_{n_i} \\ 
               =& 
    \left[
        \begin{array}{c c c}
            \begin{matrix}
                \phi^2+\tau^2 & \tau^2         & \tau^2  \\
                \tau^2        &  \phi^2+\tau^2 &  \tau^2 \\
                \tau^2        &  \tau^2        &  \phi^2+\tau^2 \\    
            \end{matrix}
            & & \bigzero 
            \\
            &
            \begin{matrix}
                \phi^2+\tau^2 & \tau^2  \\
                \tau^2        & \phi^2+\tau^2 \\
            \end{matrix}        
            & 
            \\
            \bigzero & & 
            \begin{matrix}
                \phi^2+\tau^2 & \tau^2  \\
                \tau^2        & \phi^2+\tau^2 \\
            \end{matrix} 
            \\
        \end{array}
    \right]
    \end{aligned}
\end{equation}        
\noindent where $\mathbf{I_N}$ is the identity matrix of size $N$, which is the total number of recordings, $\mathbf{1}_{n_i}$ is a matrix of ones, $N_e$ is the number of events, and $n_i$ is the number of recordings of the $i^{th}$ event.

In this approach, assuming that all recordings of the same earthquake are grouped together, the covariance matrix has a simple block diagonal form. 
The diagonal elements of $\mathbf{C}$ are equal to $\tau^2 + \phi^2$, the off-diagonal elements that are associated with recordings of the same earthquake are equal to $\tau^2$, and the remaining elements are equal to zero.

Maximum likelihood regressions are computationally inexpensive, as there are efficient methods to minimize the negative of the log-likelihood (e.g. Bound Optimization BY Quadratic Approximation (BOBYQA),  \cite{Powell2009}). 
The most involved step at each iteration is to compute the inverse of the covariance matrix; however, due to its block diagonal and sparse nature, the process is computationally efficient to perform. 

\subsection{Other Methods for Ergodic Models}

Ergodic GMMs have also been developed using Bayesian regression.
Bayesian models have been used successfully in the development of a $FAS$ GMM for Mexico City  \citep{Ordaz1994}, in deriving a $PSA$ GMM that includes the correlation between spectral periods and the correlation between the GMM coefficients \citep{Arroyo2010, Arroyo2010a}, 
in capturing the uncertainty of model parameters, such as $V_{S30}$ \citep{Kuehn2018}, and in the development of ergodic GMMs with truncated data \citep{Kuehn2020a}.
More closely related to the non-ergodic GMMs of this special issue, \cite{Hermkes2014} used a Bayesian GP regression to derive a non-parametric ergodic GMM for shallow crustal events. 
Bayesian regression has a higher computational cost than MLE which is why it is less commonly used in GMM development.

GMMs have also been derived through artificial neural networks (ANNs).
\cite{Derras2014} proposed an ANN that partitions the residuals into within-event and between-event terms and used it to develop an ergodic GMM for Europe. 
\cite{Withers2020} applied an ANN to develop an ergodic GMM with ground motions from the CyberShake simulations for Southern California.
This is a promising approach, especially for large databases, as the method scales well to many GBs of data that are frequently produced from simulation outputs ($> 10^8$ records). However, extra prudence is required as the modeler does not have direct control over the model behavior (such as interdependency among input terms) and which may limit the accurate extrapolation outside the range of training predictor variables. These concerns can be mitigated by applying physics-based constraints on the model or by augmenting the trained synthetic databases with empirical records but requires additional validation to ensure that conditions within the synthetic ground motions are consistent with empirical records and do not introduce any inherent bias within the data.

\section{Development of Partially Non-Ergodic Ground-Motion Models}

The term ``partially non-ergodic'' has sometimes been used for GMMs that include mean regional differences. 
Here, we use the term only for GMMs that include differences due to the location of the site and/or the location of the source, not for average differences between broad regions.
One such partially non-ergodic GMM approach consists of capturing systematic (i.e., site-specific) site effects \citep{Stewart2017}. 
Every site has its own velocity profile which leads to a repeatable site amplification relative to the reference profile of a GMM for the same $V_{S30}$ \citep{Lavrentiadis2021b}.
This amplification is the same for all ground motions at the site of interest and is not applicable to different sites. 
However, in an ergodic model, any misfit between a ground-motion observation and the median ground-motion estimate is considered aleatory in nature (i.e. random).
That is, ergodic GMMs are based on an assumption that the range of site amplification between different sites with the same $V_{S30}$ is the same as the range of site amplification at the site of interest. 
It is the goal of partially non-ergodic GMM to properly categorize the systematic site-specific site amplification effects and remove them from the aleatory terms.

In the partially non-ergodic model, the ergodic within-event residual is partitioned into a site-specific site term and the new remaining within-event within-site residual ($\delta WS_{es}$).
Using the \cite{AlAtik2010} notation:
\begin{equation}
    \delta W_{es} = \delta S2S_s + \delta WS_{es} 
\end{equation}

\noindent $\delta S2S_s$ term represents the systematic difference between the site amplification at the $s^{th}$ site and the site amplification in the ergodic GMM.

The parameters of a partially non-ergodic GMM can be formulated as a mixed-effects model with three random terms ($\delta B_e$, $\delta WS_{es}$, and $ \delta S2S_s$):
\begin{equation}
    y_{es} = f_{erg}(M,R,V_{S30}, ...) + \delta S2S_s + \delta B_e + \delta WS_{es} 
\end{equation}

\noindent $\delta B_e$, $\delta WS_{es}$, $ \delta S2S_s$ are assumed to be normally distributed with zero means and $\tau_0$, $\phi_{SS}$, and $\phi_{S2S}$ standard deviations, respectively. 
This leads to a more complicated covariance matrix with more non-zero off-diagonal terms:
\begin{equation} \label{eq:cov_mat_parnerg}
    \begin{aligned}
    &\mathbf{C} = \phi_{SS}^2 \mathbf{I}_N + \phi_{S2S}^2 \Sigma_{i=1}^{N_s} \mathbf{1}_{n_i} + \tau^2_0 \Sigma_{i=1}^{N_e} \mathbf{1}_{n_i} \\ 
               &= 
    \left[
        \begin{matrix}
            \phi_{SS}^2+\phi_{S2S}^2+\tau^2 & \tau^2_0                          & \phi^2_{S2S}                      & 0             \\
            \tau^2_0                        & \phi_{SS}^2+\phi_{S2S}^2+\tau^2_0 & 0                                 & \phi^2_{S2S}    \\
            \phi^2_{S2S}                    & 0                                 & \phi_{SS}^2+\phi_{S2S}^2+\tau^2_0 & \tau^2_0        \\
            0                               & \phi^2_{S2S}                      & \tau^2_0                          & \phi_{SS}^2+\phi_{S2S}^2+\tau^2_0 \\
        \end{matrix}
    \right]
    \end{aligned}
\end{equation}        

\noindent The main difference to the covariance matrix of the ergodic GMM (Equation \ref{eq:cov_mat_erg}) is that the elements that are associated with the same station include the $\phi^2_{S2S}$ variance.  
In this framework, $\sqrt{\tau^2_0 + \phi^2_{SS}}$ is the aleatory variability of the GMM, and, $\phi_{S2S}$ is the epistemic uncertainty of the site term at a site without site-specific data to constrain the site term. 

Alternatively, $\delta S2S_s$ can be estimated directly by partitioning the ergodic within-event residuals, $\delta W_{es}$, into $\delta S2S_s$ and $\delta WS_{es}$. 
This approach is expected to give similar results, but it can be problematic if some of the systematic site effects have been mapped into the ergodic event terms.

The $\delta S2S_s$ of such a partially non-ergodic GMM is spatially independent. This is a contrast with the GP-based approach  (Section \ref{sec:gp}), which allows for $\delta S2S_s$ to be spatially correlated. 

\section{Development of Non-Ergodic Ground-Motion Models}

The fully non-ergodic GMM extends the partially non-ergodic GMM to account for systematic and repeatable source and path effects in addition to the systematic site effects.
For that, two additional non-ergodic terms are added:
\begin{equation}
    y_{es} = f_{erg}(M,R_{rup},V_{S30}, ...) + \delta S2S_s + \delta P2P_{es} + \delta L2L_e + \delta B^0_e + \delta WS^0_{es} 
\end{equation}

\noindent The $\delta L2L_e$ term is the systematic source-specific adjustment to the median ground motion in the base ergodic model. 
It is related to repeatable effects in the release of seismic energy from a source in a region. 
For instance, $\delta L2L_e$ will be positive if the average stress drop of earthquakes in a region (i.e. fault system) is systematically larger than the global average.
Supporting this argument, \cite{Trugman2018} found a strong correlation between the stress drop and between-event term of an ergodic GMM.  
Similarly, the $\delta P2P_{es}$ term represents the repeatable difference in the propagation of the seismic waves between a source and site and the ergodic GMM. 
The $\delta P2P_{es}$ term will be positive if the attenuation in a geographical region is less than the global average. 

The non-ergodic terms $\delta L2L_e$ and $\delta  P2P_{es}$ are assumed to be normally distributed with zero means and $\tau_{L2L}$ and $\phi_{P2P}$ standard deviation, respectively. 
The remaining aleatory terms, $\delta B^0_e$ and $\delta WS^0_{es}$, are assumed to be normally distributed with zero means and $\tau_0$ and $\phi_0$ standard deviations.

The different GMM paradigms (e.g. ergodic, partially non-ergodic, non-ergodic GMMs) should have similar size total aleatory variability and epistemic uncertainty:
$\sqrt{\phi^2 + \tau^2} \approx \sqrt{\phi_{S2S}^2+\phi_{SS}^2+\tau^2} \approx \sqrt{\tau_{L2L}^2+\phi_{P2P}^2+\phi_{S2S}^2+\phi_0^2+\tau_0^2}$ as there is no change in the amount of information -- what is different between the three approaches is how the provided information is treated (i.e. repeatable or random). 
This is a useful check for ensuring that the epistemic uncertainty and aleatory variability of a GMM are not overestimated or underestimated. 
However, it should be noted that the size of aleatory variability and epistemic uncertainty also depends on the modeling approach. 
For example, a single-station partially non-ergodic GMM, such as SWUS15 \citep{SWUS2015} has, largely, a constant epistemic uncertainty, whereas, a non-ergodic GMM developed as GP has a scenario-dependent epistemic uncertainty. 
Therefore, this check is primarily applicable at the center of the ground motion data, not at the model extrapolation. 

\cite{Lin2011} estimated the standard deviations for all three non-ergodic terms using ground-motion data from Taiwan.
The $\delta S2S_e$ was modeled as a random term based on the site ID, and $\delta L2L_e$ was modeled as a spatially correlated random variable based on the site location using standard geostatistics.
A more complex spatial correlation model was used for $\delta P2P_{es}$, as it depends both on the source and site location. 
For a single site, the $\delta P2P_{es}$ correlation is stronger if the earthquakes are closer together, as the seismic waves travel through the same part of the crust. 
The systematic path effects were found to result in the largest reduction of the aleatory variability followed by the systematic site effects.
Overall, including all three effects led to about a $40\%$ reduction in the total aleatory standard deviation compared to the ergodic GMM.

In the previous formulation, the non-ergodic effects were modeled with normal distributions, which may not always be appropriate, particularly, for the path terms; for similar variations in the earth's crust, a far apart source-site pair will have more pronounced path effects than a source-site pair that is closer together.
The distance dependence of the path effects is not significant if all records in the dataset have similar rupture distances, but it can be important if the range of $R_{rup}$ is large.

An alternative option is to describe the non-ergodic GMM as a Varying coefficient model (VCM).
In this approach, the non-ergodic terms are scaled by different model variables (e.g. $R_{rup}$, $V_{S30}$) which provides a more flexible framework to model the systematic effects.
More details on the development of non-ergodic GMMs as GP VCMs are provided in the next section. 

\subsection{Gaussian Process Models} \label{sec:gp}

The non-ergodic GMMs in this special issue are classified as VCM, as the non-ergodic terms are dependent on the earthquake and site locations in addition to any other input parameters (e.g. $V_{S30}$):
\begin{equation} \label{eq:vcm}
\begin{aligned}
    y_{es} =& f_{nerg}(M,R_{rup},V_{S30}, ...,{t_S},{t_E}) + \delta B^0_e + \delta WS^0_{es}  \\ 
    =& f_{erg}(M,R_{rup},V_{S30}, ...) + \delta S2S(V_{S30}, ..., {t}_S) + \delta P2P(R_{rup}, ..., {t}_E,{t}_S, )  \\
    &+ \delta L2L(M, ..., {t}_E) + \delta B^0_e + \delta WS^0_{es} 
\end{aligned}
\end{equation}
\noindent with $f_{nerg}$ corresponding to the function of the median non-ergodic ground motion for a particular pair of source and site, while $f_{erg}$ is the function of the median ergodic ground motion (i.e. the median ground motion for all sources and sites).
The systematic source term, $\delta L2L$, is modeled as a function of the earthquake coordinates (${t}_E$), and the systematic site term, $\delta S2S$, is modeled as a function of the site coordinates (${t}_S$).
The systematic path term, $\delta P2P$, is more complex as it depends both on the earthquake and site location.
The cell-specific anelastic attenuation which is used to capture the systematic path effects is described in Section \ref{sec:catten}.

At first sight, the development of a non-ergodic VCM GMM may seem futile due to the large number of non-ergodic terms that need to be estimated. 
If the state of California is broken into a $5 \times 5$ km grid, there would be approximately $20{,}000$ grid points and so, at minimum, $60{,}000$ non-ergodic coefficients that would need to be estimated; that is the simplest non-ergodic model where the systematic source, site, and path effects are captured with one coefficient each. 
It is unfeasible to derive such a model with the existing datasets as they contain, at best, in the order of $10{,}000$ recordings.
Fortunately, this is not a problem in VCM due to the spatial correlation structure imposed on the non-ergodic coefficients.

In the statistical approaches described so far, the GMM coefficients are treated as fixed parameters. 
That is, every coefficient has a single value which is estimated by the MLE or another frequentist approach. 
In a GMM that is developed as a VCM GP, the model coefficients are treated as random variables that are assumed to follow Normal (Gaussian) distributions. 
The choice of the mean and covariance function of these distributions is what controls the behavior of each coefficient; for instance, whether a coefficient is constant over a domain, whether it varies continuously on some finite length scale, or whether it is spatially independent (i.e. the value of the coefficient at some location is independent of the value of the coefficient at some other location).
In this sense, in a GP regression, the GMM coefficients are modeled similarly to the aleatory terms in the mixed-effects regression (Section \ref{sec:mle}). 
It is these constraints on the GMM coefficients imposed by the covariance function that make the development of a non-ergodic VCM GP GMM tractable. 
Due to this, the non-ergodic GMM coefficients do not have to be estimated directly; instead, only the hyperparameters that control the distributions of the non-ergodic terms need to be estimated by the regression.  
With the current size of datasets, the number of hyperparameters is typically about 10. 

Furthermore, this formulation leads to a scenario-dependent epistemic uncertainty that is more appropriate than the constant epistemic uncertainty assumed in earlier studies. 
In a VCM GP GMM, the non-ergodic coefficients have a constant epistemic uncertainty, but the epistemic uncertainty of the ground motion is scaled by the GMM input variables.
For example, consider a non-ergodic GMM based on the base model (Equation \ref{eq:gmm_ask14}) where the systematic path effects are modeled with a spatially varying geometrical spreading coefficient that is a function of the earthquake coordinates ($c_{4,E}(t_E)$). 
In this case, the epistemic uncertainty of $c_{4,E}(t_E)$ will be equal to $\psi_{4,E}(t_E)$ and the epistemic uncertainty of the ground motion due to the systematic path effects will be equal to $\phi_{P2P} = \psi_{4,E}(t_E) \ln(R + c_6)$. 
This results in a distance-dependent epistemic uncertainty, the epistemic uncertainty is higher for sites farther from the source, which is different from the $\phi_{P2P}$ values of \cite{Lin2011} which are independent of the source-to-site distance. 

GP is a particular case of a hierarchical Bayesian model as it is expressed on multiple levels. At the base level are the GMM coefficients and aleatory terms which have a direct impact on the response variable $y$ and are defined in terms of some distributions; the variables that constitute this level are called model parameters ($\vec{\theta}$). 
At the next level is the set of variables that control the distributions of $\theta$. The variables of the upper level are called model hyperparameters ($\vec{\theta}_{hyp}$), which in turn could be defined in terms of some other distributions or they could be fixed. As an example, in this context, $\phi_0$ and $\tau_0$ are hyperparameters that control the distributions of the parameters: $\delta WS^0_{es}$ and $\delta B^0_e$.

There are two general approaches for developing a non-ergodic GMM with a GP regression. 
In the first approach, which was followed by \cite{Landwehr2016}, all the coefficients, ergodic and non-ergodic, were modeled as GPs. 
In that case, the non-ergodic GMM is developed from the beginning and the response variable is typically the log of the ground-motion parameter (e.g. $log(PSA)$). 
An alternative approach, which was followed by \cite{Kuehn2021} and \cite{Lavrentiadis2021}, is to model the non-ergodic coefficients or non-ergodic coefficient adjustments as GPs and keep the ergodic terms fixed. 
Here, the non-ergodic GMM is based on an existing ergodic GMM and the response variable is the ergodic residual. 
An advantage of this approach is that the extrapolation  to large magnitudes and short distances from the underlying ergodic GMM is preserved in the non-ergodic GMM.

The remaining parts of this section summarize the different elements of the VCM GP GMM development: Bayesian regression, covariance functions of the prior distributions commonly used in VCM GP GMM, cell-specific anelastic attenuation, prediction of the median, and epistemic uncertainty of the non-ergodic coefficients and median ground motion at new locations. 

\subsubsection{Bayesian Regression} \label{sec:bayes}

In Bayesian statistics, the uncertainty of the model parameters and hyperparameters, $\vec{\theta}$ and $\vec{\theta}_{hyp}$, before observing the data is expressed by the prior distribution ($p(\vec{\theta},\vec{\theta}_{hyp})$).
The uncertainty of $\vec{\theta}$ and $\vec{\theta}_{hyp}$ is updated based on the ground-motion observations, $\vec{y}$, and ground-motion parameters (such as $M$, $R_{rup}$, $V_{S30}$, etc., collectively for all records noted as $\vec{x}$) to produce the posterior distribution ($p(\vec{\theta},\vec{\theta}_{hyp}|\vec{y},\vec{x})$).
The Bayes theorem provides the means for this calculation:
\begin{equation}
    p(\vec{\theta},\vec{\theta}_{hyp}|\vec{y},\vec{x})  = \frac{ \mathcal{L}(\vec{\theta},\vec{\theta}_{hyp}) p(\vec{\theta},\vec{\theta}_{hyp}) }{p(\vec{y}, \vec{x})}
\end{equation}
\noindent Often, the normalizing distribution $p(\vec{y},\vec{x})$ is omitted for computational efficiency as it is not required to sample or compute the maximum of the posterior.
In this case the posterior distribution is expressed as: 
\begin{equation} \label{eq:bayes_prop}
    p(\vec{\theta},\vec{\theta}_{hyp}|\vec{y},\vec{x})  \propto \mathcal{L}(\vec{\theta},\vec{\theta}_{hyp}) p(\vec{\theta},\vec{\theta}_{hyp})
\end{equation}
\noindent The influence of the ground-motion data in the posterior distribution is expressed through the likelihood function ($ \mathcal{L}(\vec{\theta},\vec{\theta}_{hyp})$) --- it corresponds to the likelihood (i.e probability) of observing the data given some values for $\vec{\theta}$ and $\vec{\theta}_{hyp}$. 
The likelihood for a single observation can be estimated with the functional form of the GMM as:
\begin{equation}
    \mathcal{L}(\vec{\theta},\vec{\theta}_{hyp}) = pdf(\vec{y}|f_{nerg}(\vec{x},\vec{\theta},\vec{\theta}_{hyp}) + \delta B_e,  \phi_0^2 )
\end{equation}
\noindent Because all correlated terms (i.e. non-ergodic terms and between event residuals) are included in the mean, the misfit: $y - ( f_{nerg}(x,\vec{\theta},\vec{\theta}_{hyp}) + \delta \vec{B}_e)$, which corresponds to $\delta WS^0_{es}$, is independently and identically distributed, thus, the joint likelihood of all observations is the product of the likelihoods of individual observations:
\begin{equation} \label{eq:lik_gp}
\begin{aligned}
    \mathcal{L}(\vec{\theta},\vec{\theta}_{hyp}) &= pdf(\vec{y}|f_{nerg}(\vec{x},\vec{\theta},\vec{\theta}_{hyp}) + \delta \vec{B}_e,  \phi_0^2 )\\
    &= \Pi_{e=1}^{n_e} \Pi_{s=1}^{n_s} pdf({y_{es}}|f_{nerg}(x_{es},\vec{\theta},\vec{\theta}_{hyp}) + \delta {B}_e,  \phi_0^2 )
\end{aligned}
\end{equation}

\noindent improving computational efficiency. The likelihood function is written in vector notation in the first line and expanded in the second line of Equation \ref{eq:lik_gp}.

The prior distributions express our knowledge and beliefs about $\vec{\theta}$ and $\vec{\theta}_{hyp}$. 
They may come from prior experience in building non-ergodic GMMs or based on a desired model behavior (i.e. penalize model complexity if not supported by the data \citep{Simpson2017}).
When there is little information about $\vec{\theta}$ and $\vec{\theta}_{hyp}$, weakly informative priors can be used. 
These are chosen as wide priors distributions so that the posterior distribution is primarily controlled by the likelihood function. 

The prior distributions of the non-ergodic effects are spatially uniform with zero means and large standard deviations because prior to interrogating the ground-motion data, the systematic effects are unknown.
With the aid of the likelihood function and ground-motion data, the non-ergodic effects can be estimated close to stations and past earthquake locations.
This results in posterior distributions that are spatially varying with non-zero means and smaller standard deviations where the non-ergodic effects have been estimated.
Zero posterior standard deviations would imply that the non-ergodic effects are known with absolute certainty.

Historically, Bayesian inference has seen limited use due to its high computational cost compared to point-estimate inference with MLE.
However, in recent years, with the increase in computational speed, Bayesian models have been gaining wider adoption.
There are three main computationally-tractable approaches to obtain the posterior distributions of complex models that do not have analytical solutions; they are summarized below.  

The maximum a posteriori (MAP) approach finds the values of $\vec{\theta}$ and $\vec{\theta}_{hyp}$ that correspond to the mode of the posterior. 
The posterior distribution is proportional to the product of the likelihood function and the prior distribution (Equation \ref{eq:bayes_prop}).
MAP can be found by minimizing the negative of this product, which can be numerically computed easily with gradient-based methods. 
In this sense, MAP is equivalent to a penalized MLE where the prior distribution acts as a regularization on the likelihood function. 
MAP is computationally faster than the other numerical solutions of Bayesian models, but its main shortcoming is that it provides a point estimate not the entire posterior distribution; thus, the uncertainty of the model cannot be assessed. 
The GPML toolbox \citep{Rasmussen2010} available in Matlab and Octave provides such MAP estimates for Gaussian Process models.

The Markov Chain Monte Carlo (MCMC) approach generates samples from the posterior distributions that are used in the inference of $\vec{\theta}$ and $\vec{\theta}_{hyp}$.
This approach is able to recreate the full posterior distribution, but it is computationally slow. 
An in-depth review of this method can be found at \cite{Brooks2011}. 
Widely used statistical software that have implemented this approach are: JAGS \citep{Plummer2003}, BUGS \citep{Lunn2009}, and STAN \citep{Stan} for general Bayesian models, and GPflow \citep{GPflow2020} in Python for GPs.

A more recent approach consists in using approximation methods to compute the posterior distributions of $\vec{\theta}$ and $\vec{\theta}_{hyp}$.
These approximation solutions are applicable to specific families of Bayesian models. 
One such approximation method is the integrated nested Laplace approximation, INLA \citep{Rue2009}; it uses the Laplace approximation to efficiently compute the approximations to the marginal posterior distributions of Latent Gaussian Models (LGMs).
In this family of models, the response variable is expressed as an additive function of the model parameters ($y = \Sigma_{i=1}^n \theta_i x_i$), and all  ${\theta}_i$ follow Normal prior distributions.
INLA is a useful approximation for developing GMMs as both ergodic GMMs and non-ergodic VCM GP GMMs can be formulated as LGMs. 
Further information regarding the INLA approximation can be found in \cite{Krainski2019, Krainski2021} and \cite{Wang2018}.
A primer for developing ergodic GMMs with INLA can be found in \cite{Kuehn2021a}.

Other methods for Bayesian regression include the Variational Inference \citep{Blei2017} which approximates the posterior distribution with a member of a closed-form probability distribution.

\subsubsection{Covariance Functions} \label{sec:cov_fun}

The covariance functions, or kernel functions as often called in the literature, of the prior distributions are a crucial ingredient of the GP regression. 
They impose a correlation structure which dictates how a random variable (i.e. a coefficient or the ground-motion intensity parameter) varies in space. 
The covariance functions described in this section are isotropic and stationary; that is, the size and rate of spatial variation they impose is independent of the direction and location. 
Although this is likely a simplification for the systematic ground-motion effects, most of the non-ergodic GMM of this special issue did not use non-stationary and anisotropic kernel functions due to their additional computational challenge. 
Ground-motion studies that used non-stationary correlation structures include \cite{Kuehn2020} and \cite{Chen2021}.  
Other studies that applied non-stationary and anisotropic correlation structures to GP regressions include \cite{Paciorek2006} and \cite{Finley2011}.

The four covariance functions described here are: the identity kernel function, the spatially independent kernel function, the constant kernel function, and the exponential kernel function.
Examples, where these covariance functions are  combined to create more complex spatial correlation structures, are provided at the end of this section. 
The covariance matrices, which are used in the regression and prediction of GP, are created by evaluating the covariance functions at all indices, such as the earthquake or station IDs, or coordinate pairs, such as the earthquake or station coordinates: 
\begin{equation}
    \mathbf{K}_{i\,kl} = \bm{\kappa}_i({t}_k, {t}_l)
\end{equation}

\noindent where $\bm{\kappa}_i$ and $\mathbf{K}_i$ are the covariance function and covariance matrix for the $i^{th}$ coefficient, and ${t}_k$ and ${t}_l$ are the $k^{th}$ and $l^{th}$ indices or coordinate values. 
Indices are used as input to the kernel function if the correlation structure of the $i^{th}$ coefficient depends on information such as the event or station number, while coordinates are used as input if the correlation structure depends on information like the event or site location. 
In vector notation the covariance matrix is defined as:
\begin{equation}
    \mathbf{K}_{i} = \bm{\kappa}_i(\vec{t}, \vec{t}')
\end{equation}
\noindent where $\vec{t}$ and $\vec{t}'$ are index or coordinate arrays.
If $\mathbf{K}_{i}$ is used in the regression phase, $\vec{t}$ and $\vec{t}'$ correspond to the existing scenarios; these are the scenarios in the regression dataset. 
However, if $\mathbf{K}_{i}$ is used in the prediction phase, $\vec{t}$ and $\vec{t}'$ correspond to combinations of the existing and new scenarios.  
Further details on the GP prediction are provided in Section \ref{sec:predict}.

The identity kernel function is given by:
\begin{equation}
    \bm{\kappa}_i({t}_k, {t}_l) = \omega_{i}^2 ~ \delta(k-l)
\end{equation}
\noindent where $\delta(x)$ is the Dirac delta function ($\delta(x=0)=1$ and $\delta(x\neq0)=1$).
It is used for random variables that are statistically independent with $\omega$ being the standard deviation of the normal distribution. 
It generates a covariance matrix that is equal to $\omega^2$ along the diagonal and zero everywhere else.
This kernel function is used to model the within-event within-site aleatory term, $\delta WS_{es}$.

The spatially independent kernel function is given by:
\begin{equation}
    \bm{\kappa}_i({t}_k, {t}_l) = \omega_{i}^2 ~ \delta( \lVert {t}_k - {t}_l \lVert )
\end{equation}
\noindent This kernel imposes perfect correlation between random variables at the same location or with the same index, and zero correlation between random variables at different locations or with different indices. 
The hyper-parameter $\omega$ defines the size of the variability, that is, how much the values of the random variable vary between points that are not collocated.
If ${t}_k$ and ${t}_l$ are pair of coordinates, $\lVert {t}_k - {t}_l \lVert$ corresponds to the L2 distance norm between the two coordinates.
If ${t}_k$ and ${t}_l$ are indices, $\lVert {t}_k - {t}_l \lVert$ corresponds to the absolute difference between the two values. 

Depending on the software, the covariance matrix of a spatially independent non-ergodic term can be modeled either with the identity or spatially independent kernel function.
For example, consider a spatially independent site term, $\delta \vec{c}_{i,S}$, that has a unique value at every site but zero spatial correlation.
If a statistical software requires all terms to be of size $N$, where $N$ is the number of records, the spatially independent kernel function should be used. 
That is because, if $k^{th}$ and $l^{th}$ recording have the same station coordinates, $\delta c_{i,S\,k}$ and $\delta c_{i,S\,l}$ should be equal (i.e. perfectly correlated).
In this approach, all covariance matrices are size $N \times N$.
However, if a statistical software can model terms of different sizes, the identity kernel function can be used. 
In this case, it is more efficient to estimate $\delta \vec{c}_{i,S}$  at unique station locations and then pass it to the associated recordings. 
In this approach, $\delta \vec{c}_{i,S}$ is uncorrelated, as every station coordinate is repeated only once, thus it can be modeled with an identity covariance matrix of size $N_s \times N_s$, where $N_s$ is the number of stations, reducing the and number of operations and memory requirements.

The constant kernel function is given by:
\begin{equation}
    \bm{\kappa}_i({t}_k, {t}_l) = \omega_{i}^2 
\end{equation}
\noindent with $\omega$ controlling the deviation from the mean of the prior distribution.
It imposes perfect correlation between all random variables so that they all have the same offset from the mean function. 
As an example, in \cite{Landwehr2016}, the constant kernel function was applied to all coefficients to model their deviation from the mean of the prior which was equal to zero.

Alternatively, constant offsets in the coefficients can be modeled with a one-dimensional prior distribution on the mean function of the coefficient, as in the case of $c_{ca,p}$ in \cite{Lavrentiadis2021}. 
It depends on the modeler and software which option is more attractive. 
The main advantage of the first option is that it includes all information in the kernel function, while the main advantage of the second option is that it can lead to a sparse covariance matrix. 

The exponential kernel function is given by:
\begin{equation}
    \bm{\kappa}_i({t}_k, {t}_l) = \omega_{i}^2  ~ e^{-\frac{ \lVert {t}_k - {t}_l \lVert }{ \ell_{i}}  } 
\end{equation}
\noindent This kernel function  is applied to spatially varying random variables.
The hyperparameters $\ell$ and $\omega$ control the specific length scale and size of the spatial variation.
At the two extremes of $\ell$, the exponential kernel converges to a spatially independent and constant kernel function. 
For $\ell \rightarrow 0^+$ the spatial correlation weakens converging to a spatially independent kernel function, while for $\ell \rightarrow +\infty$ the correlation becomes stronger converging to a constant kernel function.
With this kernel function, a random variable is assumed to vary continuously but not smoothly in space (i.e. the spatial variation of the random variable is continuous, but the first derivative of the spatial variation is not). 
This kernel function is widely used in geostatistics to model spatially varying phenomena. 

Another kernel function for modeling continuously spatially varying random variables is the squared exponential.
This kernel function is infinitely differentiable resulting in very smooth spatial variations that may be unrealistic for spatial processes \citep{Stein1991}.
However, the main advantage of this covariance function is that it is separable in the $X$ and $Y$ coordinates which allows for efficient approximations of the kernel function for large datasets \citep{Lacour2022}. 

More complex correlation structures can be built by combining the kernel functions described above using the properties of the Normal distribution.
For example, assume a non-ergodic site adjustment $\delta c_{i,S}$ that is the combined effect of an underlying continuous adjustment over large distances and a site-specific adjustment. 
Such a site adjustment can be broken into individual components: $\delta c_{ia,S}$ for the underling continuous adjustment, and $\delta c_{ib,S}$  for site-specific adjustment, with $\delta c_{i,S} = \delta c_{ia,S} + \delta c_{ib,S}$.
In this case, $\delta c_{ia,S}$ can be assigned a prior distribution which has a zero prior mean and an exponential kernel function ($\bm{\kappa}_{ia,S}$), and $\delta c_{ib,S}$ can be assigned a prior distribution which has a zero prior mean and a spatially-independent kernel function ($\bm{\kappa}_{ib,S}$): 

\begin{equation}
\begin{aligned}
    \delta \vec{c}_{ia,S} &\sim \mathcal{N}\left( \vec{0}, ~ \bm{\kappa}_{ia,S}(\vec{t}_S, \vec{t}_S) \right) \\
    \delta \vec{c}_{ib,S} &\sim \mathcal{N}\left( \vec{0}, ~ \bm{\kappa}_{ib,S}(\vec{t}_S, \vec{t}_S) \right)
\end{aligned}
\end{equation}
\noindent Based on the linear properties of the Normal distribution, the prior distribution of $\delta c_{i,S}$ has a mean which is equal to the sum of mean functions of the individual components, and a kernel function which is equal to the sum of the kernel functions of the individual components:
\begin{equation}
    \delta \vec{c}_{a,S} \sim \mathcal{N}\left( \vec{0}, ~ \bm{\kappa}_{ia,S}(\vec{t}_S, \vec{t}_S) + \bm{\kappa}_{ib,S}(\vec{t}_S, \vec{t}_S) \right)
\end{equation}

Similarly, the kernel function of the median non-ergodic ground motion can be obtained by combining the kernel functions of the GMM coefficients. 
For simplicity, only three terms of the ergodic base GMM (Equation \ref{eq:gmm_ask14}) are used in this example: 
\begin{equation} \label{eq:examp_mu_fun}
    f_{nerg} = c_1 + c_{4,E}({t}_E) \ln(R_{eff}) + c_{10,S}({t}_S) \ln(V_{S30}/V_{ref})
\end{equation}

\noindent where $R_{eff} = R + c_6$, $c_1$ is the intercept, $c_{4,E}$ is the geometrical-spreading term which is a function of the earthquake coordinates and scales with $\ln(R_{eff})$, and $c_{10,S}$ is a linear site amplification term which is a function of the site coordinates and scales with $\ln(V_{S30}/V_{ref})$.

If, the GMM coefficients are modeled as GPs with prior distributions:
\begin{equation}
\begin{aligned}
    \vec{c}_1      &\sim \mathcal{N}\left( \vec{\mu}_{1},    ~ \bm{\kappa}_1 \right) \\
    \vec{c}_{4,E}  &\sim \mathcal{N}\left( \vec{\mu}_{4,E},  ~ \bm{\kappa}_{4,E}(\vec{t}_E, \vec{t}_E ) \right) \\
    \vec{c}_{10,S} &\sim \mathcal{N}\left( \vec{\mu}_{10,S}, ~ \bm{\kappa}_{10,E}(\vec{t}_S, \vec{t}_S ) \right)
\end{aligned}
\end{equation}
\noindent with $\vec{\mu}_i$ and $\bm{\kappa}_i$ being the mean and kernel functions of the $i^{th}$ coefficient, respectively; the prior distribution of $f_{nerg}$ is equal to:
\begin{equation} \label{eq:examp_mu_prior}
\begin{aligned}
    \vec{f}_{nerg} \sim \mathcal{N} \Big( & \vec{\mu}_{1} + 
                                            \vec{\mu}_{4,E} \circ \ln(\vec{R}_{eff}) + 
                                            \vec{\mu}_{10,S} \circ  \ln(\vec{V}_{S30}/V_{ref}),   \\
                                      & \bm{\kappa}_{1} + \bm{\kappa}_{4,E}(\vec{t}_E, \vec{t}_E ) \circ  (\ln(\vec{R}_{eff}) \ln(\vec{R}_{eff})^\intercal) \\ 
                                      & + \bm{\kappa}_{10,S}(\vec{t}_S, \vec{t}_S ) \circ (\ln(\vec{V}_{S30}) \ln(\vec{V}_{S30})^\intercal) \Big)
\end{aligned}
\end{equation}

\noindent in which the symbol $\circ$ corresponds to the element-wise product, and $\ln(\vec{R}_{eff})$ and $\ln(\vec{V}_{S30}/V_{ref})$ are column vectors with the $\ln(R_{eff})$ and $\ln(V_{S30}/V_{ref})$ values of all recordings. 
A linear combination of Normal distributions follows a Normal distribution.
The mean of ${f}_{nerg}$ is equal to the linear combination of the means of the prior distributions of the coefficients.
To get a more intuitive feeling for the kernel function of ${f}_{nerg}$, first consider the covariance between just two scenarios $cov({f}_{nerg\,k}, {f}_{nerg\,l})$. 
By substituting Equation \ref{eq:examp_mu_fun} into the covariance and assuming the coefficients of the GMM are independent with each other ($cov(c_i,c_j)=0$ if $i \neq j$) we obtain: 
\begin{equation} \label{eq:examp_mu_cov}
\begin{gathered}
    cov({f}_{nerg\,k}, {f}_{nerg\,l}) 
    = cov(c_1,c_1) \\
    + \ln(R_{eff\,k}) cov(c_{4,E}({t}_{E\,k}),c_{4,E}({t}_{E\,l}) ) \ln(R_{eff\,l}) \\
    + \ln(V_{S30\,k}) cov(c_{10,S}({t}_{S\,k}),c_{10,s}({t}_{S\,l}) ) \ln(V_{S30\,l}) \\ 
    = \bm{\kappa}_1 + \ln(R_{eff\,k})  \bm{\kappa}_{4,E}({t}_{E\,k},{t}_{E\,l}) \ln(R_{eff\,l}) \\
    + \ln(V_{S30\,k}) \bm{\kappa}_{10,S}({t}_{S\,k},{t}_{S\,l}) \ln(V_{S30\,l}) 
\end{gathered}
\end{equation}
\noindent The kernel function in Equation \ref{eq:examp_mu_prior} creates the same covariance as Equation \ref{eq:examp_mu_cov} for all recordings. 
For example, for the $c_{1,S}$ contribution, the $\ln(\vec{R}_{eff}) \ln(\vec{R}_{eff})^\intercal$ product creates a matrix with all $\ln(R_{eff\,k}) \ln(R_{eff\,l})$ permutations, and the element-wise product with $\bm{\kappa}_4(\vec{\chi}_e, \vec{\chi}_e )$ combines these permutations with the covariance of the coefficient. 

Generalizing from previous example, the covariance function of the median ground motion between scenarios $k$ and $l$ is:
\begin{equation} \label{eq:cov_fun}
    \bm{\kappa}_{nerg\,kl} = \Sigma_{i=1}^d ~x_{i\,k} \, \bm{\kappa}_i( {t}_{i\,k}, {t}_{i\,l}) \, x_{i\,l}
\end{equation}

\noindent in which $\bm{\kappa}_{nerg}$ is the kernel function for $f_{nerg}$, $\bm{\kappa}_i$ is the kernel function of the $i^{th}$ non-ergodic coefficient, $x_i$ is the independent variable in front of the $i^{th}$ non-ergodic coefficient (e.g. $\ln(R_{eff})$), $t_i$ is input coordinate or ID for $\bm{\kappa}_i$, and $d$ is the number of the non-ergodic terms.

In matrix notation Equation \ref{eq:cov_fun} can be defined as:
\begin{equation} 
    \bm{\kappa}_{nerg} = \Sigma_{i=1}^d \bm{\kappa}_i( \vec{t}_i, \vec{t}_i) \circ (\vec{x}_{i} \vec{x}_{i}^\intercal)
\end{equation}

\subsubsection{Cell-specific anelastic attenuation} \label{sec:catten}

\begin{figure}
    \centering
    \includegraphics[width = .40\textwidth]{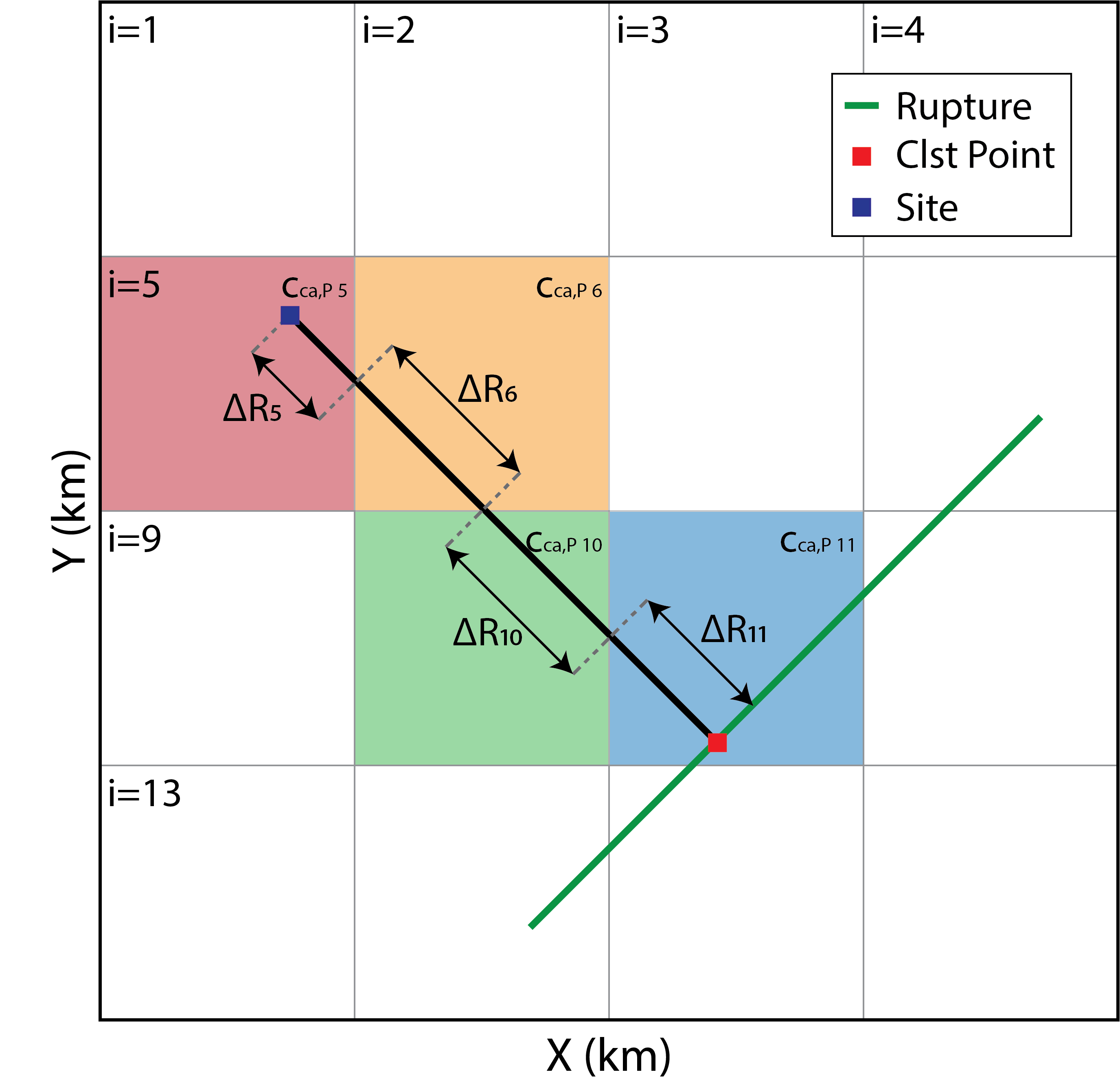}
    \caption{Schematic showing the calculation of the cell-path segments for the cell-specific anelastic attenuation. $x_{site}$ is the site location, $x_{cls}$ is the closest point on the rupture to the site, the dashed line indicates the source-to-site path, and the $\Delta R_i$ of the $i^{th}$ cell .}
    \label{fig:cells}
\end{figure}

The cell-specific anelastic attenuation was first proposed by \cite{Dawood2013} and then extended by \cite{Kuehn2019} and \cite{Abrahamson2019} as an approach to capture the systematic effects related to the paths. 
In this method, the domain of interest is divided into a grid of cells and each cell is assigned its own anelastic attenuation.
For each recording, the ray path that connects a point on the rupture with the site is broken into cell-path segments ($\Delta R_i$) which are the lengths of the ray within each cell (Figure \ref{fig:cells}).
For a given recording, the total anelastic attenuation can be calculated by $f_{atten,p} = \vec{c}_{ca,p} \cdot \Delta \vec{R}$ where $\vec{c}_{ca,p}$ is vector containing the attenuation coefficients of all the cells. 

Currently, there is no consensus on the origin point for the ray path.
\cite{Dawood2013} used the epicenter, while \cite{Kuehn2019} and \cite{Lavrentiadis2021} used the closest point on the rupture to the site, as the length of that path is equal to $R_{rup}$, which is a common distance metric for anelastic attenuation in ergodic GMMs.
Additional research is needed in this area to test different options for the origin of the ray path and also investigate if there is any magnitude dependence in the location of the representative point for finite-fault ruptures. 

In GP, the cell attenuation can be modeled similarly to the other spatially varying non-ergodic terms using a Truncated Normal as a prior distribution:
\begin{equation}
    \vec{c}_{ca,P} \sim \mathcal{N}(\vec{\mu}_{ca,P},\bm{\kappa}_{ca,P}(\vec{t}_C, \vec{t}_C)) \mathcal{T}(,0)
\end{equation}
\noindent the cell attenuation is limited to be equal or less than zero to ensure the proper extrapolation of the GMM. 
In statistical software that does not include truncated Normal distributions, the cell-specific attenuation is modeled with a Normal prior, but at a postprocessing step it is checked that no or only a small number of cells have positive attenuation.

The mean of the prior distribution, $\vec{\mu}_{ca,P}$, controls the average anelastic attenuation of the cells, while the kernel function, $\bm{\kappa}_{ca,P}$, controls their spatial correlation.
In regions with sparse coverage, the cell-specific anelastic attenuation is close to $\vec{\mu}_{ca,P}$ as there are not enough data to inform the posterior.
In regions with significant coverage, the cell-specific anelastic attenuation deviates from $\vec{\mu}_{ca,P}$ to capture the systematic path effects which influence the ground motion in those regions.
The $\vec{\mu}_{ca,P}$ can be either fixed to the anelastic attenuation of the ergodic GMM or be assigned its own prior distribution.  
The second option is computationally more involved but takes into account the re-weighting of the paths.
In an ergodic GMM, the anelastic attenuation is controlled by the attenuation of the areas with high-path coverage; however, in the cell-specific anelastic attenuation, the mean attenuation is determined at the cell level, the path coverage controls the mean and epistemic uncertainty of the individual cells, but it does not have a direct impact on the mean attenuation of all cells, which is why the mean of the cell-specific anelastic attenuation and the ergodic anelastic attenuation can be different. 

The kernel functions that were presented in Section \ref{sec:cov_fun} can also be used to model the spatial correlation of the cell attenuation. 
For instance, \cite{Kuehn2019} used the spatially independent kernel function, while \cite{Lavrentiadis2021} used a combination of the exponential and spatially independent kernel function.
Other approaches for modeling the spatial correlation of cell-specific anelastic attenuation are the conditional autoregressive (CAR) and simultaneous autoregressive (SAR) models \citep{VerHoef2018}.  
These models have sparse precision matrices (i.e. inverse of covariance matrices) reducing the computational  cost.  

The prior distribution for the total anelastic attenuation can be derived from the prior distribution for the cell attenuation using the linear transformation properties of the Normal distribution:
\begin{equation}
    \vec{f}_{atten,P} \sim \mathcal{N}( \mathbf{\Delta R} ~ \vec{\mu}_{ca,P}, \mathbf{\Delta R} ~ \bm{\kappa}_{ca,p} ~ \mathbf{\Delta R}^\intercal ) \mathcal{T}(,0)
\end{equation}
\noindent  where $\mathbf{\Delta R}$ is a matrix with the cell-path segments of all recordings, the $i^{th}$ row of $\mathbf{\Delta R}$ is equal to $\Delta \vec{R}$ of the $i^{th}$ recording. 
The $\vec{f}_{atten,P}$ prior distribution can be used in GP to make direct predictions for the median non-ergodic ground motion at new locations (Section \ref{sec:predict}).

\subsubsection{Prediction} \label{sec:predict}  

The median non-ergodic ground motion can be predicted for the new scenarios either by first predicting the non-ergodic coefficients and then substituting them at the non-ergodic functional form or by predicting the non-ergodic ground motion directly. 
This choice depends on how the GPs are modeled. 
If the GMM terms are modeled as GPs explicitly, the first method is used.
However, if the GMM terms are modeled as GPs implicitly (i.e. they have been integrated out in the likelihood function), the second method is used.  

\paragraph{Prediction of non-ergodic coefficients}

The non-ergodic coefficient adjustments for the new scenarios can be predicted based on the hyperparameters and posterior distribution of the coefficient adjustments of the existing scenarios. 
Initially, we consider the case where the non-ergodic coefficient adjustments of the existing scenarios have zero epistemic uncertainty. 
The joint prior distribution between the non-ergodic coefficient adjustments of the existing and new scenarios is:

\begin{equation} \label{eq:dc_prior}
    \left[
    \begin{matrix}
        \delta \vec{c}_i \\ \delta \vec{c}^*_i
    \end{matrix}
    \right]
    \sim
    \mathcal{N} \left( 
    \left[
    \begin{matrix}
        \vec{0} \\ \vec{0}
    \end{matrix}
    \right],
    \left[
    \begin{matrix}
    \mathbf{K}_i           & \mathbf{k}_i \\
    \mathbf{k}_i^\intercal & \mathbf{K}_i^*
    \end{matrix}
    \right]
    \right)
\end{equation}

\noindent in which $\delta c_i$ are the non-ergodic coefficient adjustments of the existing scenarios, $\delta c_i^{*}$ are the non-ergodic coefficient adjustments of new scenarios, $\mathbf{K}_i$ is the prior covariance between all pairs of existing scenarios ($\mathbf{K}_{i\,kl} = \bm{\kappa}_i({t}_k, {t}_l)$), $\mathbf{K}_i^{*}$ is the prior covariance between all pairs of new scenarios ($\mathbf{K}^*_{i\,kl} = \bm{\kappa}_i({t}^*_k, {t}^*_l)$), and $\mathbf{k}_i$ is the prior covariance between all pairs of existing and new scenarios ($\mathbf{k}_{i\,kl} = \bm{\kappa}_i({t}_k, {t}^*_l)$).

Because of the cross-correlation between the non-ergodic coefficient adjustments of the existing and new scenarios, described by $\mathbf{k}$, the posterior distributions of $\delta \vec{c}^*_i$ can be predicted by ensuring that they are in agreement with $\delta \vec{c}_i$.
A naive approach for that would be to generate multiple realizations of $\delta \vec{c}^*_i$ from the joint prior distribution (Equation \ref{eq:dc_prior}) and reject those that are inconsistent with $\delta \vec{c}_i$. 
The distribution of the accepted realizations $\delta \vec{c}^*_i$ would correspond to the posterior distribution of $\delta \vec{c}^*_i$. 
Although this is theoretically correct, it would be computationally inefficient. 
In statistics, this can be performed easily by conditioning $\delta \vec{c}^*_i$ on $\delta \vec{c}_i$, which corresponds to predicting $\delta \vec{c}^*_i$ based on the values of $\delta \vec{c}_i$.
The conditional distribution of a joint Normal prior distribution is also a Normal distribution:
\begin{equation} \label{eq:dc_post_fixed}
    \delta \vec{c}^*_i | \delta \vec{c}_i \sim \mathcal{N}(\vec{\mu}_{\delta \vec{c}^*_i|\delta \vec{c}_i}, \mathbf{\Psi}_{\delta \vec{c}^*_i|\delta \vec{c}_i} ) 
\end{equation}

\noindent with $\vec{\mu}_{\delta c^*_i|\delta c_i}$ and $\mathbf{\Psi}_{\delta c^*_i|\delta c_i}$ being the mean and covariance of the posterior distributions of $\delta c^*_i$ given by \citep{Rasmussen2006}:

\begin{equation} \label{eq:dc_mean_post_fixed}
    \mu_{\delta c^*_i|\delta c_i} = \mathbf{k}_i^\intercal \mathbf{K}_i^{-1} \delta c_i
\end{equation}
\begin{equation} \label{eq:dc_cov_post_fixed}
    \mathbf{\Psi}_{\delta c_i^*|\delta c_i} =  \mathbf{K^*}_i - \mathbf{k}_i^\intercal \mathbf{K}_i^{-1}  \mathbf{k}_i 
\end{equation}

In other fields where GP regression is used, one is typically interested only in the point-wise uncertainty which means that the mean and epistemic uncertainty of $\delta c^*_i$ can be calculated independently for each scenario reducing the computational cost. 
However, in PSHA it is necessary to calculate the full covariance for the new scenarios, as the spatial correlation of $\delta \vec{c}^*_i$, which described by the off-diagonal term of $\mathbf{\Psi}_{\delta c_i^*|\delta c_i}$, needs to be included in the logic tree. 

A more realistic case is for there to be some uncertainty in the estimation of the  non-ergodic coefficient adjustments of the existing scenarios described by the posterior distribution ($p(\delta \vec{c}_i|\vec{y},\vec{x})$). 
This uncertainty can be propagated in the prediction of the non-ergodic coefficient adjustments of the new scenarios by predicting $\delta \vec{c}^*_i$ using all possible values of $\delta \vec{c}_i$ and considering how likely each $\delta \vec{c}_i$ is ($p( \delta \vec{c}_i | \vec{y}, \vec{x} )$). 
In statistics, this is defined as marginalization of $\delta \vec{c}^*_i$:

\begin{equation} \label{eq:dc_marg}
    p(\delta \vec{c}^*_i|\vec{y},\vec{x})  = \int p(\delta \vec{c}^*_i |\delta \vec{c}_i) p( \delta \vec{c}_i | \vec{y},\vec{x} )~ d \delta \vec{c}_i 
\end{equation}

\noindent where the probability density function $p(\delta c^*_i |\delta c_i)$ can be obtained from the conditional distribution in Equation \ref{eq:dc_post_fixed}.

A closed-form solution for the posterior distribution of $\delta \vec{c}^*_i$ which includes the uncertainty of $\delta \vec{c}_i$ can be obtained if the posterior distribution of $\delta \vec{c}_i$ is assumed to be Normal \citep{Lavrentiadis2021}:
\begin{equation}
    \delta \vec{c}_i | y,x \sim \mathcal{N}(\vec{\mu}_{\delta c_i |y,x}, \mathbf{\Psi}_{\delta c_i|y,x} )
\end{equation}
\noindent where $\mu_{\delta c_i |y,x}$ is the mean, and $\mathbf{\Psi}_{\delta c_i|y,x}$ is the covariance of the posterior distribution of $\delta c_i$.
With this assumption, Equation \ref{eq:dc_marg} results in a Normal distribution:
\begin{equation}
    \delta \vec{c}^*_i | \vec{y},\vec{x} \sim \mathcal{N}(\vec{\mu}_{\delta c^*_i |y,x}, \mathbf{\Psi}_{\delta c^*_i|y,x} )
\end{equation}
with the mean and covariance given in equations \eqref{eq:dc_mean_post_marg} and \eqref{eq:dc_cov_post_marg}, respectively \citep{Bishop2006}.

\begin{equation} \label{eq:dc_mean_post_marg}
    \vec{\mu}_{\delta c^*_i|y,x} = \mathbf{k}_i^\intercal \mathbf{K}_i^{-1} \vec{\mu}_{\delta c_i|y,x}
\end{equation}
\begin{equation} \label{eq:dc_cov_post_marg}
    \mathbf{\Psi}_{\delta c_i^*|y,x} =  \mathbf{K}^*_i - \mathbf{k}_i^\intercal \mathbf{K}_i^{-1}  \mathbf{k}_i + 
        \mathbf{k}_i^\intercal \mathbf{K}_i^{-1}  \mathbf{\Psi}_{\delta c_i|y,x }  (\mathbf{k}_i^\intercal \mathbf{K}_i^{-1})^\intercal 
\end{equation}

The assumption that the posterior distribution of $\delta c_i$ is Normal is considered reasonable because in a GP regression all non-ergodic terms have Normal prior distributions. 
If the hyperparameters are fixed or follow Normal prior distributions, this assumption would be absolutely valid; however, because some of the hyperparameters are assigned different prior distributions, the posterior distribution of $\delta c_i$ may slightly deviate from this assumption. 
For the prediction of $\delta \vec{c}^*_i$, \cite{Kuehn2021} showed that this approximation gives consistent results with Equation \ref{eq:dc_marg} where the full posterior distribution of $\delta \vec{c}_i$ is used.

The non-ergodic coefficients of the new scenarios ($c^*_i$) can be computed similarly to $\delta c^*_i$, however, the non-zero prior means needs to be considered:
\begin{equation} 
    \vec{\mu}_{ c^*_i|y,x} = \mathbf{k}_i^\intercal \mathbf{K}_i^{-1} ( \vec{\mu}_{\delta c_i|y,x} - \vec{\mu}_{\delta c_i}) + \vec{\mu}_{\delta c^*_i}
\end{equation}
\begin{equation} 
    \mathbf{\Psi}_{\delta c_i^*|y,x} =  \mathbf{K}^*_i - \mathbf{k}_i^\intercal \mathbf{K}_i^{-1}  \mathbf{k}_i + 
        \mathbf{k}_i^\intercal \mathbf{K}_i^{-1}  \mathbf{\Psi}_{\delta c_i|y,x }  (\mathbf{k}_i^\intercal \mathbf{K}_i^{-1})^\intercal 
\end{equation}
\noindent where $\vec{\mu}_{\delta c_i}$ and $\vec{\mu}_{\delta c^*_i}$ are the prior means of the non-ergodic coefficients for the existing and new scenarios.  

\paragraph{Prediction of non-ergodic ground motion} 

An alternative approach to predict the median non-ergodic ground motion for the new scenarios,  $\vec{f}^*_{nerg}$, is to directly obtain it from the ground-motion observations of the exciting scenarios, $\vec{y}$ \citep{Landwehr2016}.
The main difference between this approach and the previous approach is that $\vec{y}$ includes an aleatory component which must be considered in the predictions.
The joint prior distribution between ground-motion observations of the existing scenarios and median non-ergodic ground motion of the new scenarios is:

\begin{equation}
    \left[
    \begin{matrix}
        \vec{y} \\ \vec{f}^*_{nerg}
    \end{matrix}
    \right]
    \sim
    \mathcal{N} \left( 
    \left[
    \begin{matrix}
        \vec{\mu}_f \\ \vec{\mu}^*_{f}
    \end{matrix}
    \right],
    \left[
    \begin{matrix}
    \mathbf{K}_f + \phi_0^2 \mathbf{I} + \tau_0^2 \mathbf{1} & \mathbf{k}_f \\
    \mathbf{k}^{\intercal}_f & \mathbf{K}^*_f
    \end{matrix}
    \right]
    \right)
\end{equation}

\noindent where $\mu_f$ is the prior mean of the ground-motion of the existing scenarios, and $\mu^*_f$ is the prior mean of the ground-motion of the new scenarios. 
The $\mathbf{K}_f$, $\mathbf{K}^*_f$, and $\mathbf{k}_f$ are the prior covariance for the epistemic uncertainty of the ground-motion between all pairs of existing, new, and existing/new scenarios, respectively, and $\phi_0^2 \mathbf{I}$ and $\tau_0^2 \mathbf{1}$ are the covariance for the within-event and between-event aleatory variability. 

The $\vec{\mu}_f$ and  $\vec{\mu}^*_f$ depend on how the non-ergodic GMM is being developed.
If it is based on a backbone ergodic model, $\vec{\mu}_f$ and  $\vec{\mu}^*_f$ are equal to $\vec{f}_{erg}$ for the existing and new scenarios. 
That is, without any knowledge of the non-ergodic effects, both the mean of the observations and non-ergodic ground motion are equal to the means the ergodic ground motion. 
However, if the non-ergodic GMM is developed from the beginning,  $\vec{\mu}_f$ and  $\vec{\mu}^*_f$ are equal to zero. 

The prior covariance for the epistemic uncertainty of the ground motion can be obtained by combining the kernel functions of the non-ergodic coefficients as shown in Section \ref{sec:cov_fun}, $\mathbf{K}_{f\,kl} = \Sigma_{i=1}^d {x}_{i\,k} \, \bm{\kappa}_i( {t}_{i\,k}, {t}_{i\,l}) \, {x}_{i\,l}$. 
Similarly, $\mathbf{K}^*_{f\,kl} = \Sigma_{i=1}^d {x}^*_{i\,k} \, \bm{\kappa}_i( {t}^*_{i\,k}, {t}^*_{i\,l}) \, {x}_{i\,l}^{*}$, and $\mathbf{k}_{f\,kl} = \Sigma_{i=1}^d {x}_{i\,k} \, \bm{\kappa}_i( \vec{x}_k, \vec{x}^*_l) \, {x}^*_{i\,l}$.
The prior covariance for $\vec{y}$ includes $\phi_0^2 \mathbf{I}$ and $\tau_0^2 \mathbf{1}$ because the deviation of ground-motion observations from $\vec{\mu}_y$ is the result of both aleatory variability and epistemic uncertainty. 
There is not aleatory variability in covariance between the ground-motion observations of the existing scenarios and the mean ground-motion of the new scenarios as any correlation between the two comes from the systematic non-ergodic terms. 
Similarly, the covariance of $\vec{f}^*_{nerg}$ does not include an aleatory component, as it corresponds to the median prediction of the non-ergodic ground motion. 

Once the joint prior distribution is defined, the median non-ergodic ground motion can be predicted by expressing it as a conditional distribution on the ground-motion observations:

\begin{equation} \label{eq:y_cond}
    f^*_{nerg} | y \sim \mathcal{N}(\mu_{f^*_{nerg} | y}, \mathbf{\Psi}_{f^*_{nerg} | y} ) 
\end{equation}

\noindent with the mean and the covariance of the conditional distribution given in Equations \ref{eq:y_cond_mean} and \ref{eq:y_cond_cov}.

\begin{equation} \label{eq:y_cond_mean}
    {\mu}_{f^*_{nerg} | y } = \mu^*_f + \mathbf{k}_f^\intercal (\mathbf{K}_f + \phi_0^2 \mathbf{I} + \tau_0^2 \mathbf{1})^{-1}  (y-\mu_f)
\end{equation}
\begin{equation} \label{eq:y_cond_cov}
    \mathbf{\Psi}_{f^*_{nerg} | y } =  \mathbf{K^*}_f - \mathbf{k}_f^\intercal (\mathbf{K}_f + \phi_0^2 \mathbf{I} + \tau_0^2 \mathbf{1})^{-1}  \mathbf{k}_f
\end{equation}

\subsection{Model Extrapolation Constraints and Epistemic Uncertainty}

In developing any type of GMM --- ergodic or non-ergodic --- attention must be paid to its proper extrapolation.
That is because GMMs are typically derived on datasets primarily composed of small-to-moderate earthquakes at medium-to-large distances, but in PSHA, are applied to large earthquakes at short distances. 
For that, the trends in the dataset are insufficient to guide the extrapolation of a GMM and additional constraints need to be introduced. 

These constraints can be imposed both on the model parameters as well as the model hyperparameters. 
Two common constraints for the model parameters are related to the magnitude saturation at short distances and anelastic attenuation. 

Full magnitude saturation at short distances means that, close to the fault, the ground motion does not scale with magnitude. 
Similarly, over saturation with magnitude means that, close to the fault, the ground motions reduce as the magnitude increases.
This is a controversial issue because empirical datasets, such as NGAWest2, show trends of oversaturation for large magnitudes at short periods and small distances, but the results of numerical simulations support positive magnitude scaling \citep{Collins2006, Abrahamson2007}.
Due to the limited number of empirical data from large events, and practical design purposes most GMMs do not allow oversaturation and impose full saturation as a lower limit on the regressions. 
One such practical consideration is that, if oversaturation is allowed, a structure would need to be designed not only for the largest magnitude but for the smaller events too as they could lead to higher seismic demands. 
This is straightforward to model in PSHA, but it becomes more complicated when selecting conservative deterministic scenarios. 

In a GMM, the magnitude saturation of short periods at zero distance from the rupture is controlled by the combination of the linear magnitude scaling coefficient, the geometrical spreading coefficient, the magnitude scaling coefficient for the geometrical spreading, and the pseudo-depth coefficient in geometrical spreading.
In the example GMM provided in Equation \ref{eq:gmm_ask14}, the coefficients control magnitude saturation are: $c_2$, $c_5$ and $c_6$. 
Full magnitude saturation at zero distance is achieved by:
\begin{equation}
    c_5 = \frac{-c_2}{\ln(c_6)}
\end{equation}
\noindent With this functional form, it is easy to derive a non-ergodic GMM with a full saturation constraint, as the $c_2$, $c_5$, and $c_6$ coefficients are treated as fixed terms.
However, it may more difficult to impose this constraint with other common functional forms. 
For example, \cite{ChiouYoungs2014} (CY14) uses a different functional form, and full saturation is achieved by: 
\begin{equation}
    c_2 = - c_4 ~ c_6
\end{equation}
\noindent where $c_2$ is the linear magnitude scaling, $c_4$ is the near-source geometrical spreading, and $c_6$ controls the magnitude dependence of the geometrical spreading. 
In this functional form, it is harder to include a spatially varying non-ergodic geometrical spreading as the value of $c_4$ would also affect the magnitude saturation. 
A non-ergodic GMM developer should consider factors like this when deciding on the functional form and statistical software to use.

The anelastic attenuation is intended to capture the reduction of the amplitude of the seismic waves due to the dissipation of energy as they travel through the earth's crust; thus, the anelastic attenuation coefficient or cell-specific anelastic attenuation must be negative to make physical sense. 
However, it should be noted that due to the correlation between the linear distance term and the geometrical spreading term, the physical interpretation of the linear distance term as anelastic attenuation depends on using a realistic geometrical spreading term.
Similarly to the magnitude saturation, the GMM developer should either use statistical methods and software that allows them to impose an appropriate constraint on these terms, or if that is not feasible to ensure that the model has reasonable distance scaling when used in forward calculations. 

Constraints can also be applied to hyperparameters to impose a desired model behavior. 
For instance, if a VCM GMM contains both a spatially varying site constant and a spatially varying $V_{S30}$ coefficient as a function of the site coordinates, it may be deemed reasonable to constrain the correlation length of the site constant to be smaller than the correlation length of the $V_{S30}$ coefficient.
That is because, the repeatable effects related to the site amplification due to the underlying geologic structure, which the $V_{S30}$ intends to model, are broader than the repeatable effects related to the site-specific site amplification.
Additionally, such a constraint will limit any trade-offs between the two coefficients as they would capture systematic site effects at different length scales. 

The epistemic uncertainty of a non-ergodic GMM quantifies the confidence in estimating the systematic source, path, and site effects; however, it does not quantify the confidence in the model extrapolation.
The latter is typically expressed by the model-to-model epistemic uncertainty, which reflects the range of scientifically defensible approaches for developing a GMM.
In PSHA, this uncertainty is typically captured either by using multiple GMMs or by shifting the median estimate of a base GMM.
As an example of the second approach, \cite{Abrahamson2019} incorporated the model-to-model epistemic uncertainty into a non-ergodic PSHA study for California by estimating the epistemic uncertainty and correlation of the coefficients of a common GMM functional based on the NGAWest2 GMMs and propagating them into the ground-motion prediction. 

Another source of model-to-model epistemic uncertainty for non-ergodic GMM is related to the different statistical approaches and decisions in modeling the non-ergodic terms. 
For example, different covariance functions (e.g. exponential, squared exponential) can be used to model the spatially varying non-ergodic terms or even entirely different modeling approaches (e.g. GP regression, ANNs). 
Such choices are expected to lead to bigger differences in areas with sparse data. 

Different intensity measures are affected differently by magnitude scaling. 
$PSA$, especially at short periods, is sensitive to the entire frequency content of the ground motion (i.e. spectral shape).
This can be an issue when developing a GMM predominately with small earthquakes as their frequency content is different from the frequency content of large events which are more common in PSHA, potentially resulting in incorrect scaling coefficients. 
A solution to this is developing a GMM for an intermediary intensity parameter ($IP$) that is not sensitive to spectral shape and using a transformation to convert the prediction to $PSA$.
One such example is \cite{Lavrentiadis2021c} where used $EAS$ was used as an intermediary $IP$ and Random vibration theory was used to convert $EAS$ to $PSA$.

\subsection{Other methods for non-ergodic models}

The previous sections provided an in-depth discussion on developing non-ergodic GMMs using Gaussian Process.
Although it has many useful properties, it is not the only method for developing non-ergodic GMMs.
This section provides a brief review of other methods that have been used for this task.

\cite{Sung2019} built more than 700 single-station GMMs for the Taiwan region. 
Single-station GMMs do not include non-ergodic site terms, instead, they are independently regressed with ground motions recorded at a single station. 
Kriging interpolation is used to estimate the spatial distributions of the single-station GMM coefficients and aleatory terms at new locations. 
This is a simpler approach for developing a partially non-ergodic GMM, but it cannot provide estimates of the epistemic uncertainty at the new locations as VCM GP GMM does. 

\cite{Caramenti2020} used a multi-source geographically-weighed regression (MS-GWR) to develop a non-ergodic GMM for Italy. 
It is similar to GP in that the spatial correlation of the non-ergodic terms is also captured through kernel functions; however, it is more efficient as it is based on the least-squares regression.
The main shortcoming of this approach is that the aleatory variability is described by a single term so it is unable to capture the correlation between the recordings of the same earthquake. 

\cite{Okazaki2021} developed a single-station GMM for $PGA$ using an ANN trained on strong-motion data from the KiK-net seismograph network in Japan. 
In this study, the systematic site effects were expressed as a function of site ID and estimated through the ANN fitting. 

\section{Concluding Remarks}

A summary of different methods for the development of non-ergodic GMMs is presented in this paper. 
An emphasis is placed on methods that use GP as it offers a convenient framework for expressing spatially varying non-ergodic terms. 
The cell-specific anelastic attenuation can be combined with GP to model systematic effects related to the path. 
A simple example of the steps for developing a non-ergodic GMM using a synthetic dataset and making predictions at new locations is included in the electronic supplement.

The use of non-ergodic GMMs in PSHA is a promising development, as the reduction in aleatory variability can have a large impact on the seismic hazard at large return periods, and improve the accuracy of the site-specific hazard. In PSHA applications, the reduction of the aleatory variability should be combined with the change in the epistemic uncertainty due to the uncertainty in the estimates of the non-ergodic terms in addition to the epistemic uncertainty in the extrapolation to large magnitudes and short distances of the underlying ergodic GMMs.
There is a higher computational cost associated with the development and application of non-ergodic GMMs. 
This limitation can be overcome by utilizing high-performance computers or efficient approximation methods. 
For example, INLA \citep{Rue2009} provides an efficient method for estimating the non-ergodic terms, and \cite{Lacour2019} provide an efficient approach for propagating non-ergodic effects in PSHA.
Additionally, there is an ongoing effort by the  Natural Hazards Risk and Resiliency Research Center at the Garrick Institute for the Risk Sciences at the University of California, Los Angeles to verify various software packages for developing non-ergodic GMMs which is expected to facilitate the adoption of non-ergodic GMMs. The results of that effort will be published in the near future.

As larger datasets become available, new non-ergodic GMMs are anticipated to continue adding spatially varying non-ergodic terms to capture more systematic site, path, and source effects. 
With this, non-ergodic GMMs will start to mimic the spatial resolution of numerical simulations with 3-D crustal structures. 
Numerical simulations can also be used to test the decisions and assumptions associated with non-ergodic GMM scaling \citep{Meng2021}.
There is still uncertainty in the repeatability of source effects for a given region or for a single fault.  In particular, with the use of small magnitude events to constrain the non-ergodic terms, the scaling of the non-ergodic source terms from small magnitudes to larger magnitudes has not been validated.
The variability due to fault physics complexity may inherently be irreducible at the time scales we are working with, even in consideration of fault maturity information, which is quite limited. 
In that case, non-ergodic GMMs will have a limited improvement in the accuracy of source effects. 
Similarly, path effects constrained by small events are theoretically simpler than for large events (waves emitted from different points of the fault and traversing a large volume to a site where their effect is aggregated). 
Developments in three key areas can improve non-ergodic modeling: (a) earthquake physics to help with better prediction of source effects, (b) numerical simulations to quantify the differences in path effects of large earthquakes with extended ruptures and small earthquakes with point source ruptures, and (c) continued collection of recorded motions to further constrain repeatable effects over large areas. 
Future studies should also evaluate the stability of hyperparameters between different areas to determine if a set of generic hyperparameters can be used. 
This will allow the development of non-ergodic GMM for regions with fewer recordings, as larger datasets are required to estimate the model hyperparameters than to estimate non-ergodic terms. 
Finally, even in consideration of their current limitations, non-ergodic GMMs such as those described here have advantages over ergodic (or global) GMMs in increasing the accuracy of PSHA estimates and are expected to remain a useful tool to this end.  

\section{Glossary of Proposed Notation}

\subsection{Acronyms}
{\setlength\multicolsep{0pt}
\begin{multicols}{2}
\begin{itemize}[leftmargin=1.2cm,noitemsep]
    \item [IP:]  Intensity parameter
    \item [PSA:] Pseudo spectral acceleration
    \item [FAS:] Fourier amplitude spectra
    \item [EAS:] Effective amplitude spectra
    \item [GMM:] Ground motion model
    \item [VCM:] Varying coefficient model
    \item [MLE:] Maximum likelihood estimation
    \item [GP:]  Gaussian Process
    \item [RVT:] Random vibration theory
\end{itemize}
\end{multicols}
}

\subsection{GMM Input Variables}
{\setlength\multicolsep{0pt}
\begin{multicols}{2}
\begin{itemize}[leftmargin=1.2cm,noitemsep]
    \item [$M$:]                Moment magnitude
    \item [$R_{rup}$:]          Closest distance to the rupture plane
    \item [$R_x$:]              Horizontal distance from the top of the rupture measured perpendicular to the fault strike
    \item [$R_{y0}$:]           Horizontal distance off the end of the rupture measured parallel to strike.
    \item [$\Delta\vec{R}$:]    Cell-path segments lengths of the anelastic attenuation cells cells
    \item [$V_{S30}$]           Time average shear wave velocity at the top $30 m$
    \item [$V_{ref}$]           Reference $V_{S30}$ for the linear site amplification
    \item [$z_1100$]            Depth to $1100 m/sec$ shear-wave velocity
    \item [$Dip$]               Fault dip angle.  
    \item [$F_{RV}$:]           Reverse fault scaling factor
    \item [$F_{N}$:]            Normal fault scaling factor
    \item [$f_{NL}$:]           Non-linear site amplification
    \item [$f_{HW}$:]           Hanging wall scaling
\end{itemize}
\end{multicols}
}

\subsection{Model parameters}
{\setlength\multicolsep{0pt}
\begin{multicols}{2}
\begin{itemize}[leftmargin=1.2cm,noitemsep]
    \item [$c_{i}$:]            Ergodic GMM coefficient
    \item [$c_{i,x}$:]          Non-ergodic GMM coefficient\\where $x$ can be: \\
                                $S$ for systematic site effects, \\
                                $P$ for systematic site effects, or \\
                                $E$ for systematic site source 
	\item [$\delta c_{i,x}$:]   Non-ergodic adjustment to GMM coefficient
    \item [$\vec{c}_{ca,p}$:]   Cell specific anelastic attenuation coefficients
	\item [$\delta S2S$:]       Total site-to-site non-ergodic term
    \item [$\delta P2P$:]       Total path-to-path non-ergodic term
    \item [$\delta L2L$:]       Total Source-to-source non-ergodic term
    \item [$\delta B_e$:]       Between-event aleatory term
	\item [$\delta W_{es}$:]    Within-event aleatory terms
	\item [$\delta WS_{es}$:]   Within-event within-site term of a partially non-ergodic GMM
	\item [$\delta B^0_e$:]     Between-event term of a non-ergodic GMM
	\item [$\delta WS^0_{es}$:] Within-event within-site term of a non-ergodic GMM
\end{itemize}
\end{multicols}
}

\subsection{Model hyperparameters}
{\setlength\multicolsep{0pt}
\begin{multicols}{2}
\begin{itemize}[leftmargin=1.2cm,noitemsep]
    \item [$\ell_{i,x}$:]   Correlation length in the kernel function of $c_{i,x}$ or $\delta c_{i,x}$
	\item [$\omega_{i,x}$:] Scale/Standard deviation of the $c_{i,x}$ or $\delta c_{i,x}$ kernel function
	\item [$\phi_{S2S}$:]   Standard deviation of $\delta S2S$
	\item [$\phi_{P2P}$:]   Standard deviation of $\delta P2P$
	\item [$\tau_{L2L}$:]   Standard deviation of $\delta L2L$
	\item [$\tau$:]         Standard deviation of $\delta B_{es}$
	\item [$\phi$:]         Standard deviation of $\delta W_{e,s}$
	\item [$\tau_{0}$:]     Standard deviation of $\delta B^0_{e}$
    \item [$\phi_{0}$:]     Standard deviation of $\delta {WS}^0_{e,s}$
\end{itemize}
\end{multicols}
}

\subsection{Other symbols}
{\setlength\multicolsep{0pt}
\begin{multicols}{2}
\begin{itemize}[leftmargin=1.2cm,noitemsep]
    \item [$y$:]                            Response variable of GMM
    \item [$\vec{x}$:]                      Array of GMM input variables (e.g. $R_rup$, $V_{S30}$)
    \item [$\rho$:]                         Correlation coefficient
    \item [$\vec{\theta}$:]                 Array of all GMM parameters
    \item [$\vec{\theta}_{hyp}$:]           Array of all GMM hyperparameters
	\item [$\kappa_i(\vec{t},\vec{t}')$:]   Kernel function of $c_{i,x}$ or $\delta c_{i,x}$
	\item [${t}_E$:]                        Earthquake coordinates
	\item [${t}_{Rup}$:]                    Coordinates of the closest-point on the rupture to each site
	\item [${t}_{S}$:]                      Site coordinates
	\item [${t}_{MP}$:]                     Coordinate of mid-point between source and site
	\item [${t}_{C}$:]                      Cell coordinates
	\item [$\mu(y)$:]                       Mean estimate of the $y$ ground-motion parameter
	\item [$\psi(y)$:]                      Epistemic uncertainty of $y$ ground-motion parameter
	\item [$\mu(c_i)$:]                     Mean estimate of $c_i$ coefficient
	\item [$\psi(c_i)$:]                    Epistemic uncertainty of $c_i$ coefficient
	\item [$\hat{}*$:]                      New scenarios in GP predictions (e.g. $t_E^*$ corresponds to location of new earthquake) 
	\item [$f_{erg}:$]       Median ergodic ground motion function.
	\item [$f_{nerg}:$]      Median non-ergodic ground motion function.
\end{itemize}
\end{multicols}
}

\section{Acknowledgements}
This study was supported by the Pacific Gas \& Electric Company and the California Department of Transportation. Any opinions, findings, and conclusions or recommendations
expressed in this material are those of the authors and do not necessarily reflect those of the sponsoring agencies. Constructive comments on an early draft of this manuscript provided by Linda Al Atik, Morgan P. Moschetti, Niels Landwehr, Franklin R. Olaya, and Kyle B. Withers are gratefully appreciated. 
The authors are also thankful to Jack W. Baker for the review and constructive comments that helped to improve the final article.

\section*{Declarations}
\subsection*{Funding}
Partial funding for this study has been provided by the Pacific Gas \& Electric Company and California Department of Transportation.

\subsection*{Conflict of interest}
The authors declare that they have no conflict of interest.

\subsection*{Ethics approval}
Non applicable

\subsection*{Consent to participate}
Non applicable 

\subsection*{Consent for publication}
Non applicable 

\subsection*{Availability of data and material}

\subsection*{Code availability}
A series of tools for developing and applying non-erodic VCM GP GMM are provided at:\\
\url{https://github.com/NHR3-UCLA/ngmm_tools}

\bibliographystyle{chicago}
\bibliography{references_mendeley_GL.bib, references_other.bib} 

\end{document}